\documentclass[sigconf]{acmart}
\usepackage{comment}
\usepackage{graphicx}
\usepackage{xcolor}
\usepackage[small]{subfigure} 
\usepackage[font=small]{caption}
\usepackage{balance}
\usepackage{siunitx}
\usepackage{multirow}
\usepackage{float}
\usepackage{makecell}
\usepackage{hyperref}

\AtBeginDocument{%
  }

\usepackage{ifthen}
\usepackage{amssymb}
\usepackage{xcolor}

\newboolean{showcomments}
\setboolean{showcomments}{true}

\makeatletter
\newcommand{\mynote}[3]{%
  \ifthenelse{\boolean{showcomments}}{%
   \fbox{\bfseries\sffamily\scriptsize#1}%
   {\small$\blacktriangleright$\textsf{\emph{\color{#3}{#2}}}$\blacktriangleleft$}}%
  {%
   \@bsphack
   \@esphack
  }%
}
\makeatother

\setcopyright{none}
\renewcommand\footnotetextcopyrightpermission[1]{}

\begin{document}

\title{Probabilistic Memory for Trustworthy Edge Intelligence}
\author{Likai Pei$^{*}$, Jiahao Zheng$^{*}$, Xueji Zhao$^{*}$, Emilie Ye$^{*}$, Jianbo Liu$^{*}$, Hanqing Tao$^{*}$, Ming-Yen Lee$^{\dagger }$, Ruiyang Qin$^{\ddagger }$, Yiyu Shi$^{*}$, Shimeng Yu$^{\dagger }$, X. Sharon Hu$^{*}$ and Ningyuan Cao$^{*}$}

\affiliation{%
  \institution{University of Notre Dame}
  \city{Notre Dame}
  \state{IN}
  \country{USA}
}

\affiliation{%
  \institution{Georgia Institute of Technology}
  \city{Atlanta}
  \state{GA}
  \country{USA}
}

\affiliation{%
  \institution{Villanova University}
  \city{Radnor}
  \state{PA}
  \country{USA}
}

\thanks{This paper has been accepted for publication in the proceedings of the ACM/IEEE Design Automation Conference (DAC), 2026.}
\thanks{The authors are with the College of Engineering at the University of Notre Dame, Notre Dame, IN, USA.}%





\begin{abstract}
Probabilistic computation plays an important role in trustworthy edge intelligence to quantify uncertainty, enhance robustness, reconstruct data and protect privacy, but its adoption is limited by the orders-of-magnitude data throughput gap between Gaussian random number generation (GRNG) and computation, as well as instruction overhead. This paper introduces \emph{probabilistic memory} (p-MEM), a unified memory primitive that stores distribution parameters (e.g., mean and standard deviation) and samples directly at the native memory bandwidth where deterministic data becomes the zero-variance special case. Using a layout-validated p-MEM simulator, we comprehensively explore device choices, memory specifications, and technology nodes, showing that p-MEM can achieve $>1000$\,GSa/s/mm$^2$ GRNG throughput (including memory arrays access). Integrated into CPU / GPU systems, p-MEM reduces instruction count by up to $2.19\times$/$4.37\times$, sampling latency by $562\times$/$3.45\times$, and energy by $295.5\times$/$3.53\times$ for Bayesian neural network workloads, providing a scalable hardware substrate for trustworthy probabilistic AI.
\end{abstract}



\maketitle
\pagestyle{plain}
\section{Introduction}
\label{sec:intro}

As AI systems are increasingly integrated into real-world edge platforms that directly interact with humans, their trustworthiness--including decision robustness~\cite{wangSurveyTrustworthyEdge2025}, uncertainty awareness~\cite{ICCAD}, partial observation reconstruction~\cite{wangSurveyTrustworthyEdge2025}, and privacy protection~\cite{liuPrivacybySensingTimedomainDifferentiallyPrivate2023}--has become a critical requirement. In medical domains such as wearable and implanted devices for ventricular arrhythmia detection~\cite{liu15365nmUncertaintyQuantifiable2025,abdelrazikWearableDevicesArrhythmia2025}, blood-glucose monitoring~\cite{metwallyPredictionMetabolicSubphenotypes2025}, and automated insulin delivery~\cite{diazcPerformanceBasedAdaptationIndex2025}, unreliable inference can lead to severe consequences. Similarly, in defense applications such as autonomous drone reconnaissance~\cite{yuCooperativePathPlanning2015}, robustness to sensor noise, out-of-distribution detection, reasoning under uncertainty, and protection of sensitive mission data are equally essential.

\begin{figure}[t]
  \centering
  \includegraphics[width=0.9\linewidth]{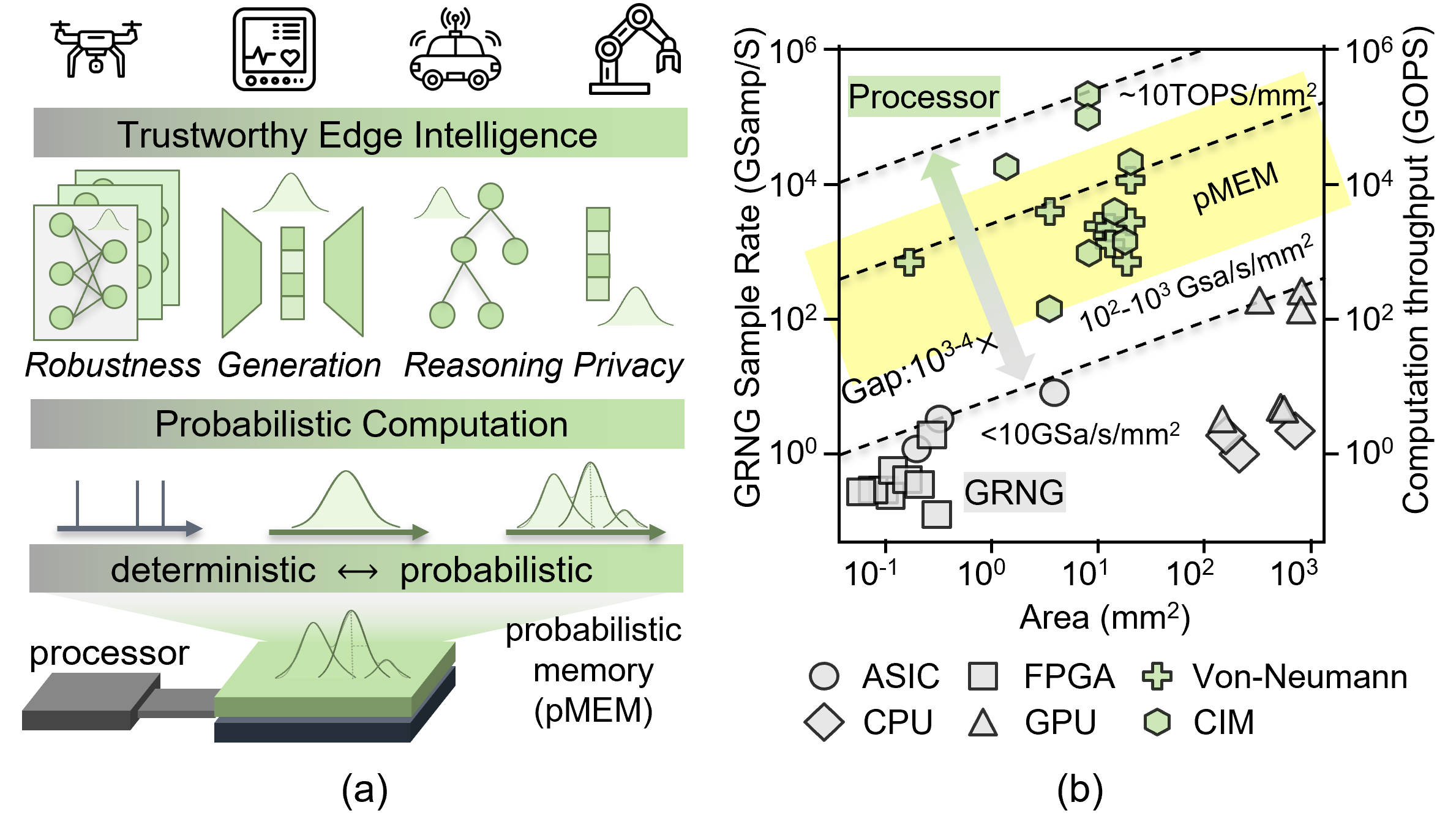}
    \vspace{-0.6cm}
  \caption{(a) Motivation for probabilistic memory to support probabilistic computation in trustworthy edge intelligence.
(b) State-of-the-art GRNG and MAC throughput across platforms and area efficiency. p-MEM is expected to close the gap required for scalable probabilistic AI.}
  \label{fig:motivation}
    \vspace{-0.6cm}
\end{figure}

To address these challenges, probabilistic computation--a foundational class of algorithms that includes Bayesian neural networks (BNNs) for robust inference~\cite{peiUncertaintyawareRoboticPerception2025}, variational autoencoders (VAEs) and diffusion models for generative reconstruction~\cite{kingmaAutoEncodingVariationalBayes2022,hoDenoisingDiffusionProbabilistic2020}, probabilistic graphical models and Bayesian decision trees for structured reasoning~\cite{kollerProbabilisticGraphicalModels2009,nutiExplainableBayesianDecision2021}, and differential privacy (DP) for data protection~\cite{dworkAlgorithmicFoundationsDifferential2014,liuPrivacybySensingTimedomainDifferentiallyPrivate2023,davisInSituPrivacyMixedSignal2024,caoStochasticMixedSignalCircuit2022a}-has emerged as a central pillar of trustworthy AI (Fig.~\ref{fig:motivation}(a)). These methods explicitly model uncertainty by sampling from learned probability distributions rather than relying on fixed parameters or deterministic operations. However, they require large-scale probabilistic sampling at high sampling intensity--the number of samples required per computation step--which far exceeds traditional uses such as weight initialization. For example, BNNs resample all weights during each inference, and differential privacy injects calibrated noise into every data vector element. This exposes a fundamental hardware bottleneck: existing random number generators (RNGs) are decoupled from the memory arrays that store distribution parameters, incurring significant latency and energy at both circuit and instruction levels. Even state-of-the-art Gaussian RNGs(GRNG) deliver less than 
10~GSa/s/mm²~\cite{liu15365nmUncertaintyQuantifiable2025,enciso65NmBayesian2025a,dorranceEnergyEfficientBayesianNeural2023a}, which is 3-4 orders of magnitude below state-of-the-art AI compute engines~\cite{shih201NVE3nm2024, qin2388836TOPSBitLevelWeightCompressed2025} (Fig.~\ref{fig:motivation}(b)). Stochastic compute-in-memory (SCIM)~\cite{yang65nmDigitalStochastic2025} leverages intrinsic device randomness~\cite{tsengReRAMBasedPseudoTrueRandom2021,piccininiSelfHeatingPhaseChangeMemoryArray2017,peiUncertaintyawareRoboticPerception2025} to accelerate probabilistic matrix–vector operations. However, probabilistic models extend well beyond dense linear algebra and often require extensive deterministic data handling (e.g., variational autoencoders, or VAE). 

This gap motivates a general-purpose probabilistic memory (p-MEM) that unifies deterministic storage and probabilistic sampling within a single memory abstraction. The main contributions of this work are summarized as follows:
\begin{itemize}
    \item Introducing a new hardware primitive--probabilistic memory--that unifies probabilistic and deterministic data handling through a novel memory-entropy decomposition and near-memory entropy integration scheme.
    
    \item Developing a cross-layer probabilistic memory simulator (\emph{pMEMSim}) to explore the throughput-density-energy trade space of p-MEM across diverse memory specifications, device technologies, and random number generation schemes, guiding scalable probabilistic memory design.
    
    \item Demonstrating up to 4.37$\times$ instruction count reduction, 546$\times$ latency reduction and 295$\times$ energy savings on representative probabilistic workloads, including Bayesian inference, probabilistic embedding search, and differential privacy.
\end{itemize}

\section{Background}
\label{sec:background}

\subsection{Probabilistic Computation}

Probabilistic computation encompasses a family of algorithms that model uncertainty by sampling from learned probability distributions during inference or data processing. BNNs~\cite{peiUncertaintyawareRoboticPerception2025} draw random samples for weights on every forward pass, resulting in extremely high sampling intensity. VAEs~\cite{kingmaAutoEncodingVariationalBayes2022} and diffusion models~\cite{hoDenoisingDiffusionProbabilistic2020} introduce stochasticity through latent-variable sampling or iterative noise injection, producing moderate but frequent sampling events. Probabilistic graphical models (PGMs)~\cite{kollerProbabilisticGraphicalModels2009} and Bayesian decision trees (BDTs)~\cite{,nutiExplainableBayesianDecision2021} rely on sampling-based inference, where randomness governs node traversal and posterior updates. DP~\cite{dworkAlgorithmicFoundationsDifferential2014,liuPrivacybySensingTimedomainDifferentiallyPrivate2023} injects calibrated noise into each data element or gradient update, yielding fine-grained, element-wise sampling. Probabilistic Cross-Modal Embeddings (PCME)~\cite{chun2021probabilistic} model semantic uncertainty directly in the representation space by mapping each image or text instance to a Gaussian distribution and computing similarity via Monte Carlo estimation. Across these methods, probabilistic sampling occurs at different granularities—per-weight, per-latent-vector, per-step, per-node, per-element, or per-embedding—and can match or exceed memory-access bandwidth, creating substantial hardware pressure on RNG throughput.
%
%

\begin{figure}[b]
\vspace{-0.4cm}
  \centering
  \includegraphics[width=0.85\linewidth]{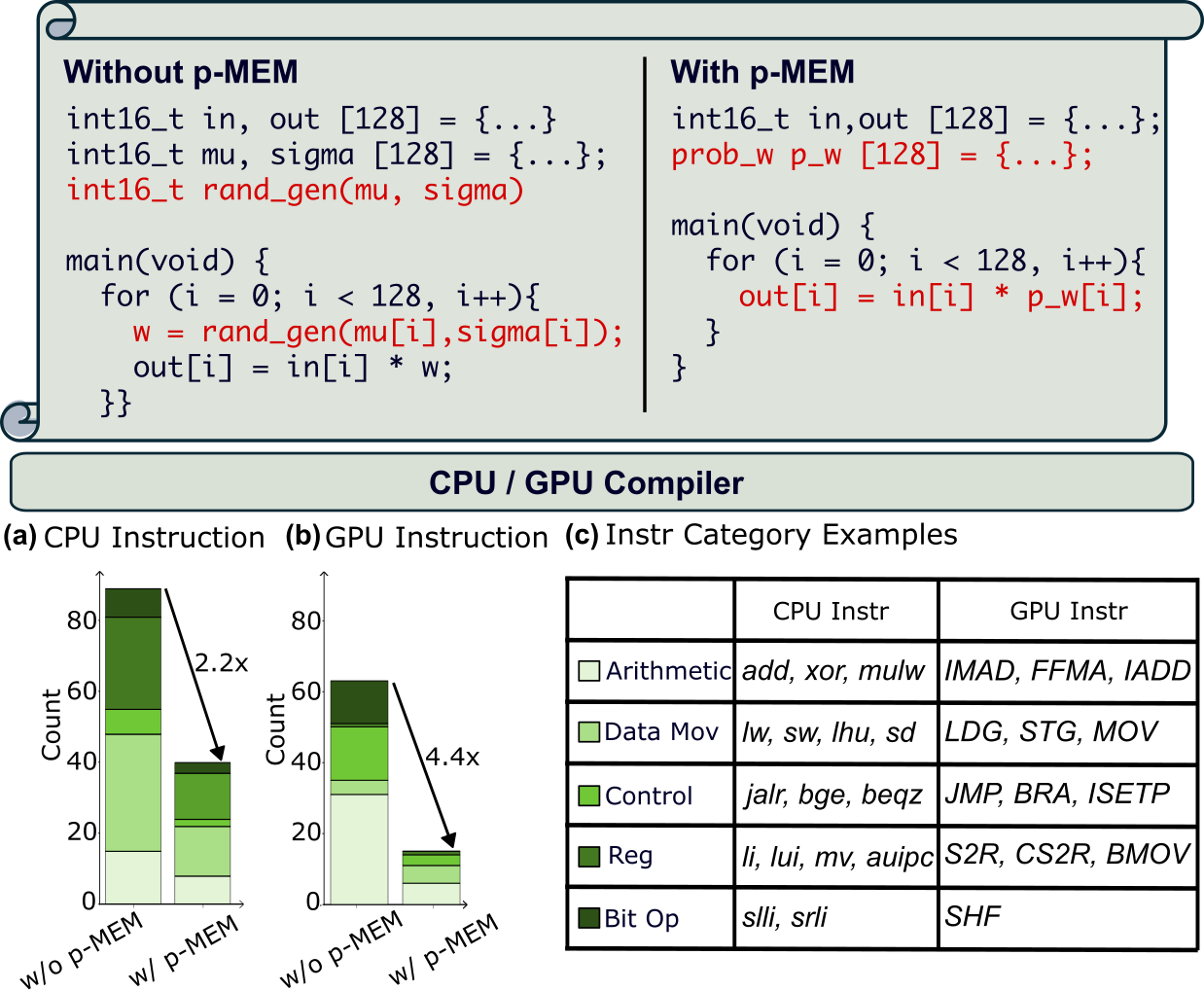}
  \caption{Compiled-instruction analysis on RISC-V CPU and NVIDIA GPU, targeting RV64GC and Ampere architecture respectively.  Instruction-level comparison demonstrating potential system-level savings when p-MEM is used as both deterministic and probabilistic memory, enabling sampling through standard memory reads. }
   \vspace{0cm}
  \label{fig:bnn_demo}
\end{figure}

\vspace{-0.2cm}
\subsection{GRNG Hardware}





Probabilistic computation fundamentally depends on high-rate sampling—typically from Gaussian distributions—making GRNG efficiency critical for hardware implementations. Digital GRNGs rely on methods such as inverse transform sampling~\cite{mullerInverseMethodGeneration1958}, central limit theorem(CLT)-based accumulation~\cite{malikRevisitingCentralLimit2015}, the Box--Muller transform~\cite{boxNoteGenerationRandom1958a}, and the Ziggurat algorithm~\cite{marsagliaZigguratMethodGenerating2000}. Even when implemented on FPGAs or ASICs, these designs achieve $<10~\mathrm{GSa/s/mm^2}$, over $100\times$ lower than state-of-the-art von Neumann processors and $1000\times$ lower than CIM-engine MAC throughput (Fig.~\ref{fig:motivation}(b)). When executed on general-purpose CPUs and GPUs, they incur additional overhead from repeated GRNG function calls. For example, without a probabilistic memory capable of sampling at native memory bandwidth and latency, vector products with probabilistic sampling---an arithmetic primitive central to BNNs and other probabilistic algorithms---require $2$--$4\times$ more instructions solely due to GRNG calls (Fig.~\ref{fig:bnn_demo}).

Analog GRNGs, on the other hand, exploit intrinsic entropy sources—such as thermal noise, supply noise, and quantum tunneling—using devices including ReRAM~\cite{tsengReRAMBasedPseudoTrueRandom2021}, PCM~\cite{piccininiSelfHeatingPhaseChangeMemoryArray2017}, CMOS circuits~\cite{parkPracticalTrueRandom2019}, and FD-SOI~\cite{peiUncertaintyawareRoboticPerception2025}. These entropy-enabled devices are often integrated into SCIM architectures~\cite{yang65nmDigitalStochastic2025}, which mandate dense vector-matrix multiplication with embedded sampling. More importantly, trustworthy-AI workloads rarely map to CIM-style dense linear algebra, as they involve heterogeneous algorithms with hybrid probabilistic-and-deterministic data access. For example, BDT follows sequential sampling paths, while VAEs exhibit partial stochasticity—sampling only latent variables while relying on deterministic encoder and decoder networks. This mismatch motivates a general-purpose memory that unifies deterministic and stochastic data access, providing native, flexible, and scalable support for the diverse probabilistic algorithms essential to trustworthy AI.

\begin{figure*}[t]
    \vspace{-0.5cm}
    \centering
    \includegraphics[width=0.85\textwidth]{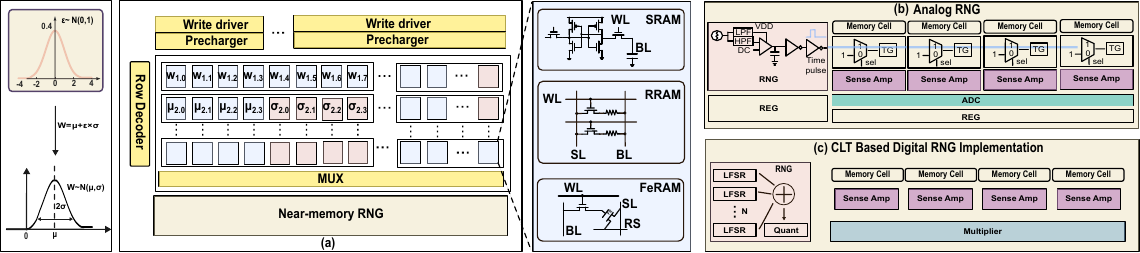}
    \caption{Overall architecture of the proposed p-MEM. (a) Weight decomposition and hybrid deterministic / probabilistic data storage. (b) Near-memory analog RNG circuits. (c) Near-memory digital RNG circuits.}
    \label{fig:architecture}
    \vspace{-0.5cm}
\end{figure*}

\vspace{-0.2cm}
\section{Probabilistic Memory Architecture}

\subsection{Design Requirement and Capabilities}

To provide the required functionalities for trustworthy probabilistic computation, an ideal p-MEM should exhibit the following key capabilities:
(1) It supports both deterministic data and probabilistic parameters at arbitrary memory addresses using standard memory-write operations. 
(2) It provides unified reads of deterministic and probabilistic data. In the probabilistic mode, p-MEM directly samples from the stored distribution, enabling single-instruction sampling and a unified load interface for both data types, which substantially reduces instruction count in probabilistic computation (Fig.~\ref{fig:bnn_demo}). 
(3) The added sampling functionality incurs minimal area overhead, preserves memory density, and does not increase the energy or latency of deterministic reads and writes. 
(4) The architecture is device-agnostic---e.g., compatible with 6T SRAM, memristors, FeFETs, and other emerging devices---and supports diverse randomness sources (analog noise or digital logic), multiple memory configurations (e.g., MUX ratio and array size), and scalability to advanced technology nodes and multiple memory hierarchies. Below, we introduce our p-MEM design to achieve these capabilities.

\vspace{-0.2cm}
\subsection{p-MEM Architecture}
Probabilistic data following a Gaussian distribution $w_{ij} \sim \mathcal{N}(\mu, \sigma)$ are commonly decomposed as $w_{ij} = \mu + \epsilon \cdot \sigma$, where $\epsilon \sim \mathcal{N}(0,1)$ is generated at runtime, as demonstrated in the compute-in-entropy (CIE) architecture~\cite{ICCAD}. This separates each stochastic weight into two deterministic parameters, $\mu$ and $\sigma$, stored in memory, and a dynamic stochastic component, $\epsilon$. By storing $(\mu,\sigma)$ and generating $\epsilon$ near the memory, p-MEM enables high-throughput sampling while avoiding repeated weight transfers.

Building on this, p-MEM introduces a unified memory architecture where $(\mu,\sigma)$ are stored conventionally, and $\epsilon$ is generated near-memory. Each cell supports two modes: deterministic (returns stored value) and probabilistic (computes $w = \mu + \epsilon \cdot \sigma$). This design removes the need for explicit GRNG instructions and supports sampling through the standard memory read path.

Because the architecture is device-agnostic, it applies to both CMOS-based SRAM and emerging devices such as RRAM, FeRAM. 
It is also compatible with both analog and digital $\epsilon$-generation schemes.

\begin{table}[b]
\caption{Comparison of statistical test results for different CLT numbers in digital GRNG versus analog RNG implementation.}
\vspace{-0.3cm}
\centering
\small
\renewcommand{\arraystretch}{1.25}
\setlength{\tabcolsep}{6pt}
\resizebox{0.9\columnwidth}{!}{
\begin{tabular}{c|cccccc|cc}
\toprule
\multirow{2}{*}{\textbf{Metric}} 
& \multicolumn{6}{c|}{\textbf{Digital GRNG (CLT Number)}} 
& \multicolumn{2}{c}{\textbf{Analog GRNG}} \\ 
\cmidrule(lr){2-9}
& \textbf{2} & \textbf{4} & \textbf{6} & \textbf{8} & \textbf{12} & \textbf{16} & \cite{ICCAD} & \cite{peiUncertaintyawareRoboticPerception2025}\\
\midrule
\textbf{KS Test $p$-value} 
& 0.002 & 0.21 & 0.48 & 0.504 & 0.694 & 0.728 & 0.591 & 0.757 \\
\textbf{$\chi^2$ Test $p$-value} 
& 0 & 0 & 0 & 0.003 & 0.141 & 0.29 & 0.137 & 0.707\\
\midrule
\textbf{Equivalent CLT Level}
& \multicolumn{6}{c|}{\textcolor{blue}{Analog $\approx$ CLT = 12--16}} & -- & -- \\
\bottomrule
\end{tabular}}
\label{tab:grng_clt_comparison}
\end{table}

\vspace{-0.2cm}
\subsection{Near-Memory Randomness Generation}
To generate the stochastic variable $\epsilon$ with minimal latency and data movement, p-MEM supports two forms of near-memory RNG: analog RNG and digital RNG.

\noindent \textbf{Analog RNG:}
In the analog configuration (Fig.\ref{fig:architecture} (b)), every four columns share a common RNG and ADC to balance area efficiency and read throughputwlarger sharing ratios reduce area but also limit parallelism. Based on this trade-off and prior findings\cite{ICCAD} showing that 4-bit $\sigma$ precision provides sufficient uncertainty information, we adopt 4-bit as the minimum granularity for probabilistic units. Higher-precision formats (e.g., 8-bit or 16-bit) can be constructed by grouping multiple 4-bit units. Each cell includes a small multiplexer and a mode-control register to support both deterministic and probabilistic modes.

As shown in Fig.~\ref{fig:architecture} (b), we sample the supply voltage and extract both low- and high-frequency noise components using low-pass and high-pass filters, respectively~\cite{ICCAD}. The resulting signals are amplified through an operational amplifier, then sampled by a capacitor. The sampled voltage is converted into a pulse via an inverter, which is used to control the charge and discharge behavior of the bitline.

In the probabilistic mode, the cell receives stochastic pulse streams from the shared RNG. These stochastic excitations modulate the charge-discharge dynamics of the cell, and the resulting analog waveform is subsequently digitized by an ADC.
In the deterministic mode, the stochastic path is bypassed entirely, and the stored value is accessed directly through a sense amplifier. 
A compact register stores the sign bit associated with the stochastic output.

\noindent \textbf{Digital RNG:}
In the digital configuration, each group of four probabilistic cells shares a dedicated digital RNG. 
This RNG uses a CLT-based scheme that forms pseudo-Gaussian noise by summing uniform random variables generated by configurable 16-bit linear-feedback shift registers (LFSRs). 
The number of LFSRs per block is tunable to adjust RNG quality. 
Each block also integrates a local multiplier to compute the $\sigma \epsilon$ product in the digital domain before contributing to the final sampled output.

\noindent \textbf{Quality Metrics and Comparison:}
The quality of digital RNGs, based on the CLT, depends on the number of accumulated uniform variables. To determine a suitable CLT depth and ensure fair comparison with analog designs, we evaluate output distributions using two statistical metrics: the Kolmogorov-Smirnov (KS) test~\cite{masseyjrKolmogorovSmirnovTestGoodness1951} and the Chi-square ($\chi^2$) test~\cite{nikulinChiSquaredGoodnessofFitTest2013}, both reported through p-values.

As shown in Table.\ref{tab:grng_clt_comparison}, increasing the number of uniform sources in CLT-based digital RNGs improves statistical quality, indicated by higher p-values in KS and Chi-square tests. The two analog RNGs achieve comparable or better results than digital CLT implementations with more than 12 accumulations. Therefore, CLT-12 is selected as the standard digital configuration for fair comparison.

\vspace{-0.2cm}
\section{Design Space Exploration}
\label{sec:system}


\subsection{P-MEM Simulator Overview }

\begin{figure}[t]
  \vspace{-1cm}
  \centering
  \includegraphics[width=0.78\linewidth]{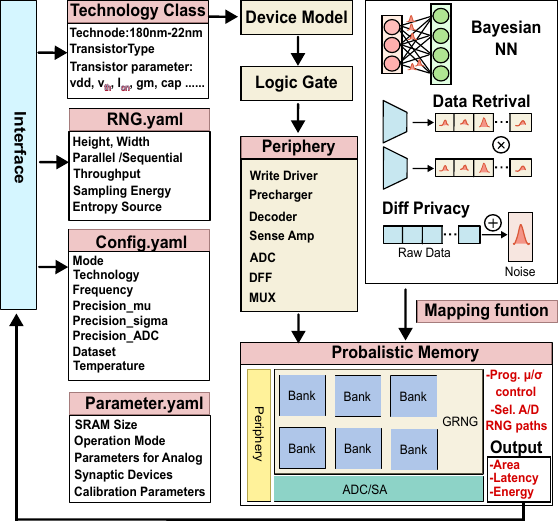}
  \caption{Framework overview of proposed p-MEM.}
  \label{fig:tool_diagram}
  \vspace{-0.7cm}
\end{figure}

The backbone of our probabilistic memory simulator, \emph{pMEMSim}, is based on NeuralSim~\cite{luNeuroSimSimulatorComputeinMemory2021}, which originally models deterministic CIM architectures. To model p-MEM systems, we retain its memory and peripheral modeling infrastructure while extending it to incorporate integrated RNGs and stochastic behavior, as shown in Fig.\ref{fig:tool_diagram}. The framework consists of an interactive interface, configurable parameters file structure, and a simulation engine.

The interface allows users to configure modeling parameters such as memory type, bit precision, $\mu$/$\sigma$ ratio, application workload, supply voltage, clock frequency, and ADC precision. It also supports importing user-defined device and RNG models. All parameters are organized into structured files and passed to the simulation engine.

The simulation engine models the full probabilistic memory stack hierarchically—from devices to logic gates, peripheral blocks, and the system level. SRAM device models are based on Predictive Technology Models (PTM)~\cite{PTM}, supporting nodes from 180nm to 22nm. In contrast, RRAM and FeFET models are derived from fabricated data provided in the NeuroSim framework. The technology file defines electrical parameters, the config file specifies $\mu$ and $\sigma$ bit precision and system-level settings, and the parameter file contains customizable physical attributes such as capacitance and resistance. RNG specifications follow a fixed input format to support diverse implementations via circuit-level metrics.

From the device models, we construct a logic gate library (INV, NAND, NOR), capturing key parameters like area, capacitance, resistance, and drive strength. Higher-level circuits—such as WL/column decoders, sense amplifiers, ADCs, prechargers, write drivers, and level shifters—are built from these primitives. The p-MEM model supports hardware-level configuration of analog and digital RNG integration. Disabling the $\sigma$ path yields deterministic behavior; additional implementation details are discussed in later sections.

At the system level, application-specific configurations generate a workload mapping file that defines operation modes and memory access patterns. The simulator reports three key metrics: area, latency, and energy. Area is modeled with layout-aware constraints and placement rules. Latency is estimated using the Horowitz model~\cite{nvsim}, considering gate and interconnect delay. Dynamic energy is calculated using E=C$V^2$, including parasitic capacitance.


\vspace{-0.2cm}
\subsection{Validation}

Our framework\footnote{\url{https://github.com/CSIRLab/PROMISE}} builds on NeuroSim by reusing some circuit parameters and structures, but modifies the overall architecture and peripherals to support probabilistic computation.

We performed layout-level validation of key building blocks under a 65nm technology node to evaluate the accuracy of area estimation. As shown in Table. \ref{tab:eda_pred_comparison}, the simulated results for representative modules are compared against extracted layout data. The average relative error across modules is approximately 3.05$\%$, demonstrating good agreement between our model and actual layout implementations.

For latency and energy validation, we benchmarked the analytical results from our simulator against detailed SPICE-level simulations using Cadence Virtuoso. The average error in latency estimation is 13.87$\%$, and for energy estimation the average error is 14.48$\%$ across multiple modules and configurations. The latency of the DFF module is simplified in our framework and is assumed to be equal to the system clock cycle for modeling convenience.

\vspace{-0.2cm}
\subsection{Performance Evaluation on 65nm CMOS Memory}

\begin{table}[t]
\caption{Comparison of predicted vs. EDA-based results for Area, Latency, and Energy.}
\vspace{-0.3cm}
\centering
\small
\setlength{\tabcolsep}{3pt} %
\renewcommand{\arraystretch}{1.2} %
\resizebox{0.9\columnwidth}{!}{
\begin{tabular}{c|ccc|ccc|ccc}
\toprule
\multirow{2}{*}{\textbf{Module}} 
& \multicolumn{3}{c|}{\textbf{Area ($um^{2}$)}} 
& \multicolumn{3}{c|}{\textbf{Latency (ps)}} 
& \multicolumn{3}{c}{\textbf{Energy (fJ)}} \\ 
\cmidrule(lr){2-4} \cmidrule(lr){5-7} \cmidrule(lr){8-10}
& \textbf{EDA Tool} & \textbf{Predicted} & \textbf{Error} 
& \textbf{EDA Tool} & \textbf{Predicted} & \textbf{Error} 
& \textbf{EDA Tool} & \textbf{Predicted} & \textbf{Error} \\
\midrule
MUX               &2.675   &2.697  & 0.8\%   &42.8  &50  & 16.8\%  &0.976 &1.17  & 19.8\% \\
DFF               &12.76   &12.856  & 0.75\%  &-  &-  & -\%  &4.275  &4.217  & 1.3\% \\
Level Shifter     &20.52   &19.95  & 2.7\%  &40.96  &47.44  & 15.8\% &206.5  &240.4  &14.1 \% \\
Precharger        & 7.0626 & 7.012 & 0.7\% & 347.4 & 291.8 & 16\% &99.48   &113.7 &14.29  \% \\
Sense Amp         & 34.42 & 32.5 & 5.57\% & 359.9& 335 & 6.9\%  & 18.66 & 14.37 &  22.9\% \\
\bottomrule
\end{tabular}}

\label{tab:eda_pred_comparison}
\vspace{-0.4cm}
\end{table}

\noindent\textbf{Evaluation Metrics:} For performance evaluation at the subarray level, we define several key metrics. Area per bit is used to quantify the storage efficiency of the memory. Latency refers to the time required for a single read operation, while read dynamic energy quantifies the energy consumed per access. In addition, we define bandwidth per area, expressed in bit/s/mm², as a measure of memory throughput efficiency under different architectural configurations.

\begin{figure}[b]
\vspace{-0.3cm}
    \centering
    \includegraphics[width=0.4\textwidth]{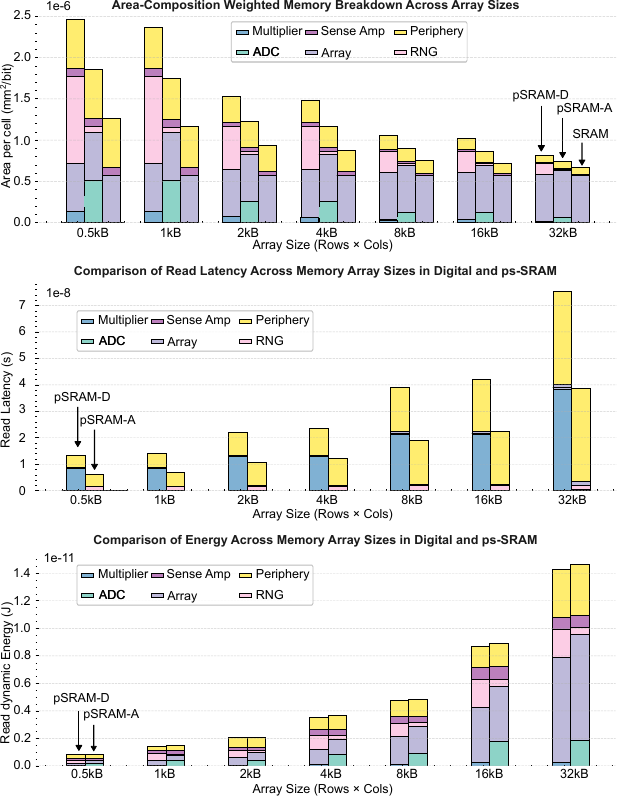}
    \caption{Comparison of pSRAM-D, pSRAM-A, and conventional SRAM under different array sizes in terms of area per cell, read latency, and dynamic energy. Each bar indicates the proportional contribution of different module components.}
    \label{fig:ps_analog_digital_result1}
    \vspace{-0.5cm}
\end{figure}

\noindent\textbf{Comparative Evaluation:}
Fig.~\ref{fig:ps_analog_digital_result1} presents a comparison of the area per bit among three memory architectures: SRAM, pSRAM-D (digital, CLT-based RNG), and pSRAM-A (analog, noise-based RNG), under a configuration with a CLT size of 12, a 4-bit ADC, and a MUX-sharing ratio of 8. Conventional SRAM exhibits the smallest area per bit, whereas pSRAM-A and pSRAM-D incur 6.1$\%$/21.2$\%$ degradation, respectively, at a 32~kB array size. The digital design shows 1.95× higher latency than the analog version, mainly due to the use of multipliers in the $\sigma\epsilon$ computation. While the analog design achieves faster stochastic integration, it consumes slightly more energy due to ADC usage in each read. The difference is minimal, with only a 2.2$\%$ energy overhead compared to the digital design.

\begin{figure}[t]
  \centering
  \includegraphics[width=0.85\linewidth]{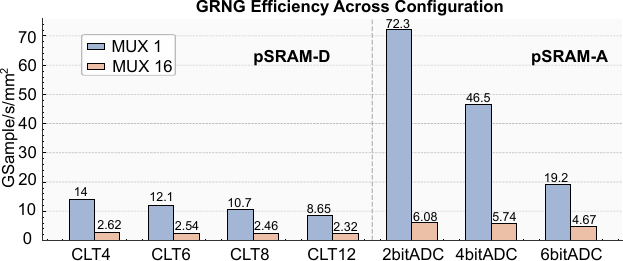}
  \caption{Bandwidth-per-area comparison between pSRAM-D and pSRAM-A under different configurations.}
  \label{fig:tradeoff2}
  \vspace{-0.5cm}
\end{figure}

\noindent\textbf{GRNG Throughput per Area:} 
Fig. \ref{fig:tradeoff2} presents the GRNG throughput per area under various configurations for both analog and digital implementations. For the analog design, different ADC precisions are explored, while for the digital counterpart, we sweep across multiple CLT accumulation depths. In addition, several MUX sharing ratios are evaluated to understand their impact on performance. The figure shows results based on a 128$\times$128 array, corresponding to a 2kB memory instance. Across the evaluated configurations, the peak bandwidth per area reaches up to 72.3Gsamples/s/mm², effectively closing the performance gap observed in Fig. \ref{fig:motivation} and demonstrating the scalability of the proposed probabilistic memory architecture under high-throughput design targets.

\begin{figure}[b]
\vspace{-0.5cm}
    \centering
    \includegraphics[width=0.38\textwidth]{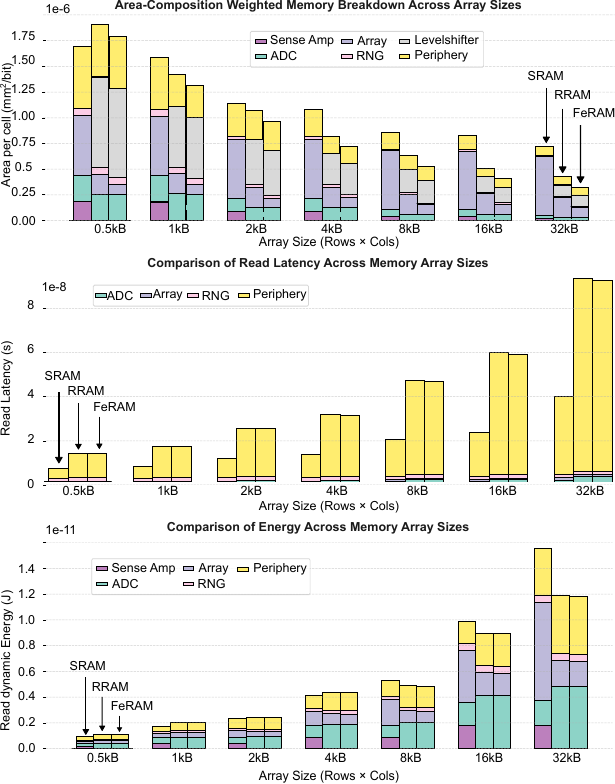}
    \caption{Comparison of SRAM, RRAM, and FeRAM in area, latency, and energy across array sizes, with module-wise breakdown.}
    \label{fig:ps_sram_rram_result}
    \vspace{-0.5cm}
\end{figure}

\subsection{Cross-Technology and Memory-Type Evaluation}

Fig.~\ref{fig:ps_sram_rram_result} compares three memory cell types—SRAM, RRAM, and FeRAM under the pMEM-Analog architecture across various array sizes, with a 4-bit ADC and a MUX sharing ratio of 8. At a 32~kB array size, FeRAM and RRAM demonstrate significant area efficiency improvements over SRAM, reducing per-bit area by 54$\%$ and 39.7$\%$, respectively, due to their higher cell densities and simpler peripheral requirements. In terms of latency and energy, nonvolatile memories exhibit similar trends, with both RRAM and FeRAM incurring approximately 2.3× higher latency compared to SRAM. However, they achieve energy savings of up to 18.87$\%$, making them attractive options for energy-constrained applications despite the access speed trade-off.

\begin{figure}[t]
    \centering
    \includegraphics[width=0.4\textwidth]{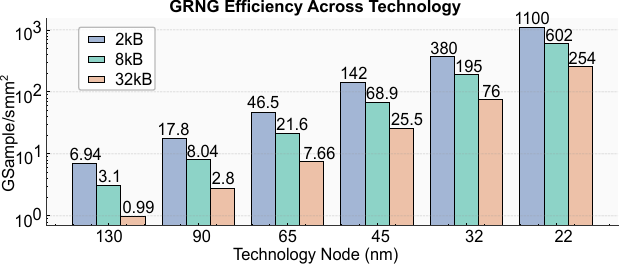}
    \vspace{-0.2cm}
    \caption{Bandwidth per area comparison across different array sizes and technology nodes.}
    \label{fig:BW_cross_technology}
    \vspace{-0.8cm}
\end{figure}

As shown in Fig.\ref{fig:BW_cross_technology}, the bandwidth per area of pSRAM-A, configured with (4-bit ADC, 8 MUX ratio), is evaluated across different array sizes and technology nodes. Since these results also take into account memory array (not only near-memory RNG), smaller SRAM array show better GRNG area efficiency. Results show consistent improvement with scaling, reaching up to 1100GSa/s/mm² at the 22nm node.

\section{P-MEM Benchmarking}
\label{sec:evaluation}

\subsection{Algorithm Evaluation}


\noindent\textbf{Bayesian Neural Network:}
BNNs are distinguished by their ability to provide probabilistic estimates of uncertainty in predictions. 
Unlike classical neural networks, which optimize for a single best set of parameters, BNNs infer a posterior distribution over weights, thereby capturing multiple plausible solutions to a given task. 
Formally, the posterior distribution after observing the dataset \(D\) is given by Bayes’ rule: $p(\mathbf{w}|D) = \frac{p(D|\mathbf{w}) \, p(\mathbf{w})}{p(D)}$, where \(p(D|\mathbf{w})\) denotes the likelihood of the data given the weights, and \(p(\mathbf{w})\) is the prior distribution over the weights. 



For BNNs, which are designed to capture predictive uncertainty beyond deterministic neural networks, algorithm-level evaluation focuses on classification accuracy and uncertainty calibration. To assess the latter, we use Expected Calibration Error (ECE), a standard metric that quantifies the discrepancy between a model’s predicted confidence and its actual accuracy. A lower ECE indicates better-calibrated uncertainty estimates.

\noindent\textbf{Differential Privacy:}
Differential privacy (DP)\cite{liuPrivacybySensingTimedomainDifferentiallyPrivate2023} provides a rigorous framework for protecting individual information in data analysis. 
A randomized mechanism $\mathcal{A}: D \to \mathbb{R}^d$ satisfies $(\varepsilon,\delta)$-DP if, for any two neighboring datasets $D \sim D'$ differing in a single record and for any set $S \subseteq \mathbb{R}$, \(\Pr[\mathcal{A}(D) \in S] \leq e^{\varepsilon} \Pr[\mathcal{A}(D') \in S] + \delta\), 
Here, $\varepsilon > 0$ bounds the privacy loss, while $\delta \in [0,1]$ allows a small probability of failure. 
The Gaussian mechanism~\cite{dworkAlgorithmicFoundationsDifferential2014,kasiviswanathanWhatCanWe2010} achieves $(\varepsilon,\delta)$-DP by injecting noise sampled from the normal distribution, calibrated to the $\ell_2$-sensitivity of the query. 
For a function $f : D \to \mathbb{R}^d$, the mechanism is defined as $\mathcal{A}_G(D; f, \varepsilon, \delta) = f(D) + (Y_1, Y_2, \ldots, Y_d)$, where $Y_i \sim \mathcal{N}(0,\sigma^2)$ for $i=1,\ldots,d$. 
A common calibration is $\sigma = \frac{\Delta^2_f \sqrt{2 \ln(1.25/\delta)}}{\varepsilon}$, with $\Delta^2_f$ denoting the $\ell_2$-sensitivity of $f$.

\begin{table}[h]
\caption{Comparison of full precision and quantized (8-bit $\mu$, 4-bit $\sigma$) settings across applications.}
\vspace{-0.3cm}
    \centering
    \resizebox{\columnwidth}{!}{
    \begin{tabular}{cccccc}
        \hline
        Application & Dataset & Precision Config & ECE (\%) ↓ & ACC (\%) ↑  \\
        \hline
        \multirow{2}{*}{BNN} & \multirow{2}{*}{CIFAR-10}
            & Full Precision            & 3.1 & 88.93  \\
            &                          & 8-bit $\mu$, 4-bit $\sigma$ & 3.7 & 88.91  \\
        \hline
        \multicolumn{2}{c}{ } & Precision Config & V→T R@1 (\%) ↑ & V→T R@10 (\%) ↑ \\
        \hline
        Deterministic   & MSR-VTT 
            & Full Precision            & 26.7  & 72.5 \\
        \multirow{2}{*}{PCME} & \multirow{2}{*}{MSR-VTT}
            & Full Precision            & 37.2 & 73.7 \\
            &                          & 8-bit $\mu$, 4-bit $\sigma$ & 37.1 & 73.8 \\
        \hline
    \end{tabular}}
    \label{table:multi_app_quant_new}
    \vspace{-0.5cm}
\end{table}

\noindent\textbf{Probabilistic Cross-Modal Embeddings:}
Probabilistic embedding approaches model \emph{semantic uncertainty} 
directly in the representation space. In cross-modal retrieval, where a single 
image may correspond to multiple textual descriptions (and vice versa), this 
ambiguity is naturally expressed by representing each sample as a probability 
distribution rather than a deterministic vector. 

PCME~\cite{chun2021probabilistic} exemplifies this approach by mapping each image or 
text instance to a multivariate Gaussian $\mathcal{N}(\boldsymbol{\mu}, 
\mathrm{diag}(\boldsymbol{\sigma}^2))$ in a shared latent space, where 
$\boldsymbol{\mu} \in \mathbb{R}^D$ encodes the canonical embedding and 
$\boldsymbol{\sigma}^2 \in \mathbb{R}^D$ quantifies sample-specific semantic 
uncertainty. To compute similarity between two samples, PCME estimates the 
expected matching probability via reparameterized sampling. Given 
$\mathbf{z}_i \sim \mathcal{N}(\boldsymbol{\mu}_i, \mathrm{diag}(\boldsymbol{\sigma}_i^2))$ 
and $\mathbf{z}_j \sim \mathcal{N}(\boldsymbol{\mu}_j, \mathrm{diag}(\boldsymbol{\sigma}_j^2))$, 
the similarity is defined as
\( p_{ij} = \mathbb{E}_{\mathbf{z}_i, \mathbf{z}_j}\!\left[\sigma(\alpha - \beta \|\mathbf{z}_i - \mathbf{z}_j\|^{2})\right] \), 
where $\sigma(\cdot)$ denotes the sigmoid function, and $\alpha, \beta$ are 
learnable temperature parameters. In practice, this expectation is approximated 
via Monte Carlo sampling with 10-100 samples per pair.


For probabilistic cross-modal embeddings, which model uncertainty in video-text matching, we evaluate both retrieval accuracy and computational efficiency. We use standard recall metrics (R@1) to measure retrieval performance.


\noindent\textbf{Dataset and Network Architecture: }
To evaluate the BNN algorithm in the context of image classification, we adopt the CIFAR-10~\cite{cifar10} dataset and use MobileNet~\cite{mobilenet} as the backbone. CIFAR-10 is a widely used benchmark dataset containing 60,000 color images across 10 classes. MobileNet is designed for low-power devices with depthwise separable convolutions that significantly reduce computation and model size, making it well-suited for edge deployment.


To evaluate probabilistic cross-modal embeddings, we use the MSR-VTT dataset~\cite{xu2016msrvtt}, a large-scale video-text retrieval benchmark with 10,000 video clips and paired textual descriptions. Following standard practice, we train on the official training split (6,513 videos) and evaluate on the 1kA test set—1,000 curated video-text pairs for consistent benchmarking. We adopt ImageBind-Huge~\cite{girdhar2023imagebind} as a frozen backbone to extract 1024-dimensional embeddings for both modalities. Our probabilistic projector is a two-layer MLP (2048 hidden units) that maps embeddings to Gaussian parameters (mean and log-variance), enabling uncertainty-aware similarity via Monte Carlo sampling (10 samples per inference).

\noindent\textbf{Evaluation Results: }
As shown in Table~\ref{table:multi_app_quant_new}, we evaluate the performance of deterministic and probabilistic models across both BNN and cross-modal retrieval tasks. To align with our hardware configuration, $\mu$ is quantized to 8 bits and $\sigma$ to 4 bits. Results indicate that probabilistic models consistently outperform their deterministic counterparts, demonstrating the benefits of uncertainty modeling. Moreover, low-precision quantization introduces minimal performance degradation.

\vspace{-0.3cm}
\subsection{System Benchmark}
To systematically evaluate the benefits of pMEM, we construct a benchmark framework based on three representative algorithms. Each exhibits a different interaction pattern between deterministic and probabilistic data. In the BNN benchmark, the core computation involves deterministic inputs multiplied by probabilistic weights. In the PCME scenario, the operation consists of element-wise multiplication between two probabilistic vectors. For DP inference, deterministic inputs are combined with additive probabilistic noise.

For each algorithm, we implement the core computation in C and compile (RISCV-GCC) to obtain the baseline instruction count. In this flow, pMEM is used to replace instructions related to random number generation and memory access, effectively offloading the sampling workload into the memory fabric. Energy and latency are further estimated via hardware-level simulation on multiple platforms, including CPU (Intel Xeon W-2265) and GPU (NVIDIA RTX A5000). The p-MEM configurations are tailored to each workload. For BNN and DP, we adopt a 4-4-2kB architecture--comprising four mats, each with four subarrays. PCME uses custom 16-4-32kB pMEM configurations optimized for their respective access patterns. As shown in Table. \ref{table:platform_comparison_reordered}, p-MEM integration leads to significant reductions in instruction count and improvements in both latency and energy per operation. Specifically, on CPU/GPU platform, for BNN, we observe a $2.19\times$/$4.2\times$ reduction in instructions, along with $562\times$/$3.45\times$ speedup and $295.5\times$/$3.53\times$ energy savings. The savings come from p-MEM generated random weight, avoiding the software random sampling and associated computation and data movement. PCME achieves a $1.67\times$/$4.37\times$ gain in instruction count, $546\times$/$2.95\times$ latency reduction, and $270\times$/$2.89\times$ lower energy consumption. DP inference similarly benefits from a $2.19\times$/$4.2\times$ instruction reduction and $46\times$/$2.27\times$ improvements in latency and $35.6\times$/$2.25\times$ improvements in energy.

\begin{table}[t]
\caption{Performance comparison of different hardware configurations across applications. Latency and energy are reported as the average over $10^5$ vector–vector operations. }
\vspace{-0.5cm}
\centering
\resizebox{\columnwidth}{!}{
\begin{tabular}{ccccc}
\hline
Application & Platform & Instruction Count (M) & Latency (us/op) ↓ & Energy (uJ/op) ↓ \\
\hline

\multirow{4}{*}{BNN}
    & CPU             & 33.7  & 40.81  & 2600.90   \\
    & CPU + p-MEM      & 15.37 ($\times$2.19) & 0.072 ($\times$562.8) & 8.801 ($\times$295.5)  \\
    & GPU             & 23.6  & 0.11   & 8.99        \\
    & GPU + p-MEM      & 5.62 ($\times$4.2) & 0.034 ($\times$3.45) & 2.543 ($\times$3.53) \\
\midrule

\multirow{4}{*}{PCME}
    & CPU             & 701.7  & 40.2303     & 2976.04   \\
    & CPU + p-MEM      & 419.8 ($\times$1.67) & 0.073 ($\times$546.6) & 10.99 ($\times$270.7)  \\
    & GPU             & 671.7  & 0.1123    & 8.0951    \\
    & GPU + p-MEM      & 153.6 ($\times$4.37) & 0.038($\times$2.95) & 2.7986 ($\times$2.89)  \\
\midrule

\multirow{4}{*}{DP}
    & CPU             & 0.069   & 41.371       & 2673.723     \\
    & CPU + p-MEM      & 31.48 ($\times$2.19) & 0.898 ($\times$46) & 75.073 ($\times$35.61)  \\
    & GPU             & 48.38   & 0.112    & 8.339    \\
    & GPU + p-MEM      & 0.011 ($\times$4.2) & 0.0495($\times$2.27) & 3.6987 ($\times$2.25)  \\
\hline

\end{tabular}}
\vspace{-0.4cm}
\label{table:platform_comparison_reordered}
\end{table}

\color{black}

\vspace{-0.3cm}
\section{Conclusion}
\label{sec:conc}

\vspace{-0.1cm}
We propose a general, scalable, and programmable p-MEM architecture that bridges the performance gap between RNG sampling and memory access. To support comprehensive design exploration, we develop a full-stack simulation framework spanning device, circuit, and system levels. The framework enables parametric analysis of key configurations such as ADC precision, MUX ratio, and RNG types. We validate p-MEM across three representative algorithms--BNN, PCME, and DP--on both CPU and GPU platforms. Experimental results show that p-MEM reduces instruction count by $2.19\times$/$4.2\times$, achieves up to $546.6\times$/$3.6\times$ latency improvement, and delivers $295.5\times$/$3.65\times$ energy savings, demonstrating its effectiveness for efficient and trustworthy probabilistic computing.

\vspace{-0.3cm}
\section{Acknowledgments}

This work was supported by the National Science Foundation under Grant No. 2404874.

\clearpage
\bibliographystyle{unsrt}
\bibliography{references,datasets,reference1}
\balance
\end{document}